\pdfoutput=1
\documentclass[12pt]{article}

\usepackage{amsmath,graphicx}
\usepackage{epsf}
\usepackage{graphicx,epsfig}
\usepackage{amsfonts}
\usepackage{amssymb}
\usepackage[nosort]{cite}
\usepackage{setspace}

\def\bk{{\bf k}}
\def\bp{{\bf p}}
\def\bq{{\bf q}}

\def\CO{{\cal O}}

\def\mpl{M_{\rm P}}
\def\half{\frac{1}{2}}

\makeatletter
\renewcommand\section{\@startsection {section}{1}{\z@}%
                                 {-3.5ex \@plus -1ex \@minus -.2ex}%
                                   {2.3ex \@plus.2ex}%
                                   {\normalfont\large\bfseries}}
\renewcommand\subsection{\@startsection{subsection}{2}{\z@}%
                                   {-3.25ex\@plus -1ex \@minus -.2ex}%
                                     {1.5ex \@plus .2ex}%
                                     {\normalfont\bfseries}}
\renewcommand\subsubsection{\@startsection{subsubsection}{3}{\z@}%
                                   {-3.25ex\@plus -1ex \@minus -.2ex}%
                                     {1.5ex \@plus .2ex}%
                                     {\normalfont\itshape}}
\makeatother

\newcommand{\Letter}{
\setlength{\textwidth}{16.5cm}
   \setlength{\textheight}{22.6cm}
    \hoffset=-0.5in
\voffset=-2.1cm }

\Letter

\begin{document}
\newcommand{\be}{\begin{equation}}
\newcommand{\ee}{\end{equation}}
\newcommand{\bea}{\begin{eqnarray}}
\newcommand{\eea}{\end{eqnarray}}
\newcommand{\barr}{\begin{array}}
\newcommand{\earr}{\end{array}}
\def\bal#1\eal{\begin{align}#1\end{align}}

\thispagestyle{empty}
\begin{flushright}
\end{flushright}

\vspace*{0.3in}
\begin{spacing}{1.1}

\begin{center}
{\large \bf Relic Vector Field and CMB Large Scale Anomalies}

\vspace*{0.5in} {Xingang Chen$^{1,2}$ and Yi Wang$^{2,3}$}
\\[.3in]
{\em
$^1$Department of Physics, The University of Texas at Dallas, Richardson, TX 75083, USA\\
$^2$Centre for Theoretical Cosmology, DAMTP,
\\University of Cambridge, Cambridge CB3 0WA, UK\\
$^3$Kavli Institute for the Physics and Mathematics of the Universe (WPI),
\\Todai Institutes for Advanced Study, University of Tokyo,
\\5-1-5 Kashiwanoha, Kashiwa, Chiba 277-8583, Japan
} \\[0.3in]
\end{center}

\begin{center}
{\bf
Abstract}
\end{center}
\noindent
We study the most general effects of relic vector fields on the inflationary background and density perturbations. Such effects are observable if the number of inflationary e-folds is close to the minimum requirement to solve the horizon problem. We show that this can potentially explain two CMB large scale anomalies: the quadrupole-octopole alignment and the quadrupole power suppression. We discuss its effect on the parity anomaly. We also provide analytical template for more detailed data comparison.

\vfill

\newpage
\setcounter{page}{1}


\newpage

\section{Introduction}
\setcounter{equation}{0}

Inflation was introduced to solve the horizon, flatness, homogeneity, isotropy and relic problems in the initial conditions of the Big Bang model \cite{Guth:1980zm}.
The fluctuations generated by inflation are nearly scale invariant \cite{s2}, which provides initial conditions for the cosmic microwave background (CMB) and the large scale structure (LSS).
If we are lucky and the inflation only lasted for the minimal number of e-folds that is necessary to solve these problems, we may be able to see traces of the initial conditions at the largest scales in the sky. The presence of various large scale anomalies in the CMB \cite{Bennett:2010jb,Ade:2013nlj} suggests that this could indeed be the case.

For example, after subtracting off the Doppler dipole, the planes of the first two CMB multipole components, the quadrupole and octopole, are anomalously aligned \cite{Tegmark:2003ve}. It is also observed that, up to $\ell = 22$, the CMB multipole seems to be anomalously parity-odd, with the powers $C_{\rm odd} > C_{\rm even}$ \cite{Land:2005jq, Ade:2013nlj}. In addition, the power of the quadrupole is particularly low \cite{Hinshaw:1996ut}.
To assess how likely these anomalies have cosmological origin and study how much we may learn about the physics in the initial moments of the inflation, we need both careful investigations in data analyses and theoretical model building for inflation.
See \cite{Copi:2010na} for a review.

Among various possible primordial relics, vector fields are natural sources for anisotropy and they exist ubiquitously.\footnote{For examples of other types of relic anisotropies, see \cite{Pereira:2007yy}.} They are present in Standard Model, and they can also be present in any sectors that are coupled to or hidden from the inflaton sector.
To study the most general effects of these fields, in our model, we turn off any direct coupling between the vector field and the inflaton.
So the vector fields get diluted during inflation, which is their most natural fate.\footnote{To maintain the field strength of the vector field during inflation, one has to introduce special types of couplings between the vector field and the inflaton \cite{Turner:1987bw}.} Initially their density can be high and this is what we will study in this paper.
We study the impact of these vector fields on the inflation background and the inflationary density perturbations. We show that these effects may be simultaneously responsible for two of the anomalies mentioned above, namely the quadrupole-octopole alignment and low quadrupole. Individually, each of these anomalies has only modest statistical significance. If they indeed have cosmological origins, perhaps the only way to improve our understanding is to have interpretations that can account for multiple of them. For this purpose, we also present analytical template suitable for more detailed data analyses.
We also discuss the effect on the parity anomaly, which cannot be successfully addressed by the present results.

\section{Background}
\setcounter{equation}{0}

We start with the action
\bea
S=\int d^4x \sqrt{-g}
\left[ \frac{\mpl^2}{2} R - \frac{1}{4} F_{\mu\nu} F^{\mu\nu}
- \half \partial_\mu \phi \partial^\mu \phi - V(\phi) \right] ~.
\eea
We will set $\mpl=1$ except at the final results.
At the leading order, $V(\phi) = V_0$ is a constant and the inflaton $\phi$ drives the inflation.
We place the vector field along the $z$-direction,
\bea
F_{03}=-F_{30}=E(t) ~,
\eea
and the other components of $F_{\mu\nu}$ are zero.
This vector field back-reacts on the gravity and induces anisotropy to the metric.

Due to the axial symmetry and spatial translational invariance, the metric takes the following form
\bea
ds^2 = -N(t) dt^2 + 2C(t) dt dz + A^2(t) (dx^2+ dy^2) + B^2(t) dz^2 ~.
\eea
We can redefine $t$ and $z$ to make $N(t)=1$ and $C(t)=0$.
The equations of motion are
\begin{align}
\partial_\mu \left( \sqrt{-g} F^{\mu\nu} \right) &=0 ~,
\label{EOM_F}
\\
R_{\mu\nu} - \half g_{\mu\nu} R &= F_{\rho\mu} F^\rho_\nu
- \frac{1}{4} g_{\mu\nu} F_{\rho\sigma} F^{\rho\sigma} - g_{\mu\nu} V_0 ~.
\label{EOM_G}
\end{align}
The equation (\ref{EOM_F}) tells us that
\bea
E(t) = \frac{B}{A^2} E_0 ~,
\eea
where $E_0$ is a constant.
The Einstein equations (\ref{EOM_G}) become
\begin{align}
H_A^2 + 2 H_A H_B &= \frac{E_0^2}{2 A^4} + V_0 ~,
\\
\dot H_A + \dot H_B + H_A^2 + H_B^2 + H_A H_B &=
-\frac{E_0^2}{2A^4} + V_0 ~,
\\
2 \dot H_A + 3 H_A^2 &= \frac{E_0^2}{2A^4} + V_0 ~,
\end{align}
where we have defined
\bea
H_A \equiv \frac{\dot A}{A} ~, \quad
H_B \equiv \frac{\dot B}{B} ~.
\eea
We can integrate these equations and get the following exact solution,
\begin{align}
A(t) &= a \sqrt{1+ \frac{E_0^2}{8 a^4 H_0^2}} ~,
\label{solution_A}
\\
B(t) &= a \frac{1-\frac{E_0^2}{8 a^4 H_0^2}}{\sqrt{1+ \frac{E_0^2}{8 a^4 H_0^2}}} ~,
\label{solution_B}
\end{align}
where $a=\exp (H_0 t)$ is the attractor scale factor and $H_0=\sqrt{V_0/3}$ is a constant. We have assumed that the vector field is the only source of anisotropy and chosen the integration constant so that, in the limit of $E_0\to 0$, we recover the isotropic inflationary solution with the scale factor $a$.

This exact solution describes a Bianchi type I universe filled with an electric field and a cosmological constant \cite{Jacobs}. At the point $t_b = \frac{1}{2H_0} \ln \frac{E_0}{2\sqrt{2}H_0}$, the scale factor in the $z$-direction is zero, while those in the $x$ and $y$-direction are finite, supported by the vector fluxes. As $t>t_b$, the vector field and the anisotropy get diluted, and the universe approaches the attractor dS space.
In this paper we shall focus on the $t>t_b$ region and do not make reference to the $t\leq t_b$ epoch.

In the $t>t_b$ region, the electric field takes value
$E_0^2 < 8 a^4 H_0^2$
in Planck units. For example, for large field inflation, we shall have $E_0/a_0^2 \lesssim (10^{16}\mbox{GeV})^2$, where $a_0$ is the scale factor evaluated at an initial time $t_0 > t_b$, which shall be canceled when calculating the physical quantities such as energy density of the electric field. In order for the effect from the electric field to be observationally significant, we shall require $E_0$ to be close to this upper bound value.

We treat the kinetic term of the inflaton as a probe to this background.
For simplicity, we assume that, at the limit $E_0\to 0$, the inflaton evolves according to the usual attractor solution determined by the constant slow-roll potential slope $\partial_\phi V$.\footnote{More generally, the inflaton may not be in the attractor yet, since this is the beginning of the inflation.}
The evolution of $\phi$ can also be solved exactly,
\bal
\dot\phi = \dot\phi_0  \frac{1 + \mathbb{E}}{1-\mathbb{E}} ~,
\qquad \mathbb{E}\equiv \frac{E_0^2}{8 H_0^2 a^4} ~,
\label{dotphi}
\eal
where $\dot\phi_0 = -\partial_\phi V/(3H_0)$ is the inflaton velocity in the absence of the vector field.

\section{Density perturbations}
\setcounter{equation}{0}

Now we consider the inflationary density perturbation on this background. We study the quantum fluctuations of the inflaton field $\phi$.
We expand the perturbation of the inflaton as $\phi(x)=\phi_0(t)+\delta\phi(x)$, and quantize
\begin{align}
  \delta\phi(x) = \int \frac{d^3 k}{(2\pi)^3}
  \left[
    u_\mathbf{k} a_\mathbf{k} + u^*_{-\mathbf{k}} a_{-\mathbf{k}}^\dagger
  \right]  e^{i \mathbf{k}\mathbf{x}} ~,
\end{align}
with the usual creation and annihilation operators, $a_\bk$ and $a_{-\bk}^\dagger$.
The equation of motion for $u_\mathbf{k}$ can be written as
\begin{align} \label{eq:eom-nonpert}
  u''_\mathbf{k} + \frac{2(1+\mathbb{E}+2\mathbb{E}^2)}{\tau(1-\mathbb{E}^2)} u'_\mathbf{k}
  + \left[
    \frac{k_x^2+k_y^2}{1+\mathbb{E}} + \frac{(1+\mathbb{E})k_z^2}{(1-\mathbb{E})^2}
  \right]  u_\mathbf{k}=0 ~,
\end{align}
where the prime denotes the derivative with respect to the conformal time $\tau \approx -1/(aH)$.
The significance of the impact of the vector field on the background is measured by the parameter $E_0^2/(8 H_0^2 a^4)$, which we will simply denote as $\CO(E_0^2)$. To obtain analytical solution for $u_k$, we work in the perturbative regime in which this parameter is small. Such an epoch always exists because the scale factor $a$ is expanding, and it is close to the beginning of the inflationary spacetime.
We can perturbatively solve \eqref{eq:eom-nonpert} order by order in $E_0^2$ using
\bal \label{eq:pert-eom-a1}
\frac{1}{a^2} \frac{d}{d\tau} \left( a^2 \frac{d}{d\tau} u_{\bk(1)} \right)
+ k^2 u_{\bk(1)}
=
\frac{1}{a^2} \frac{d}{d\tau}
\left( \frac{E_0^2}{16 a^2 H_0^2} \frac{d}{d\tau} u_{\bk(0)} \right)
+ \frac{E_0^2}{ 16 a^4H_0^2} (3k_x^2 + 3k_y^2 - 5k_z^2) u_{\bk(0)}
~,
\nonumber\\
\eal
where we have denoted the different orders in $u_\bk$ as
\bea
u_\bk = u_{\bk(0)} + u_{\bk(1)} +\cdots ~.
\eea
We get, up to order $E_0^2$,
\bea
u_\bk = C_+ u_{\bk(0)}
+ \frac{H_0^3}{\sqrt{2k^3}} E_0^2 \sum_{n=3}^6 \alpha_{+n} \tau^n e^{-ik\tau}
~,
\label{u_expand}
\eea
where
\begin{align}
u_{\bk(0)} &= \frac{H}{\sqrt{2k^3}} (1+ik\tau) e^{-ik\tau} ~,
\\
\alpha_{+3} &= -\frac{i}{24 k^3} (3k_x^2 + 3k_y^2-4k_z^2) ~,
\\
\alpha_{+4} &= \frac{1}{24 k^2} (3k_x^2 + 3k_y^2-4k_z^2) ~,
\\
\alpha_{+5} &= \frac{i}{40 k} (3k_x^2 + 3k_y^2-4k_z^2) ~,
\\
\alpha_{+6} &= \frac{1}{80} (-k_x^2 - k_y^2 +3k_z^2) ~.
\end{align}
This perturbative solution is valid for $|k\tau| \ll \mathbb{E}^{-1} \sim H_0^{-2} E_0^{-2} \tau^{-4}$, which is sufficient for the purpose of this paper because we are mostly interested in modes that are close to the horizon.
Here we only considered the solution with a positive frequency, which we define as the Bunch-Davies (BD) component. We will discuss more general case in the next section.
The $C_+$ is a time-independent normalization constant, determined by the canonical quantization condition between $\delta\phi$ and its momentum conjugate $\delta\pi$,
\bea \label{eq:quant}
[ \delta \phi_\bp, \delta \pi_\bq ] = i (2\pi)^3 \delta^3(\bp+\bq) ~,
\eea
which is
\bea
a^3 ( 1-\frac{E_0^2}{16 a^4H_0^2} )
(u_\bk \dot u_\bk^* -  u_\bk^* \dot u_\bk) = i ~.
\label{W-cond}
\eea
To $\CO(E_0^2)$,
this condition gives
\begin{align}
|C_+|^2
&= 1 -\frac{E_0^2 H_0^2}{8 k^6} (3k_x^2 + 3k_y^2-4k_z^2)
\nonumber \\
&= 1- \frac{1}{8}(3-7 \cos^2\theta) \frac{E_0^2 H_0^2}{k^4}
~,
\end{align}
where $\theta$ is the angle between the comoving momentum $\bk$ and the direction of the vector field ${\bf E_0}$.

Using the time delay formula $\zeta \approx -H \delta\phi/\dot\phi$, and evaluating the background parameters with their attractor values, we get the power spectrum
\bal
\langle \zeta^2 \rangle = \frac{P_\zeta}{2 k_1^3} (2\pi)^5 \delta(\bk_1+\bk_2) ~,
\eal
with
\bal
P_\zeta = P_{\zeta0}
\left( 1 - \frac{3 E_0^2 H_0^2}{8\mpl^2 k^4} + \frac{7 E_0^2 H_0^2}{8 \mpl^2 k^4} \cos^2\theta \right)~,
\label{power_BD}
\eal
where $P_{\zeta0} = H^4/(2\pi\dot\phi_0)^2$ is the power spectrum in absence of the vector field. Note that here the most important part of the power spectrum (\ref{power_BD}) is the anisotropic term with strong scale dependence $\propto \cos^2\theta/k^4$.
It comes from the quantum fluctuation $\delta\phi$. All the other background parameters only affect the isotropic part of the result.
We also comment that although when evaluating the power spectrum, the second term in (\ref{u_expand}) vanish after taking the late time $\tau \to 0$ limit, it plays an important role determining the coefficient $C_+$ through (\ref{W-cond}).

\section{The non-Bunch-Davies case}
\setcounter{equation}{0}

\begin{figure}[t]
  \centering
  \includegraphics[width=0.47\textwidth]{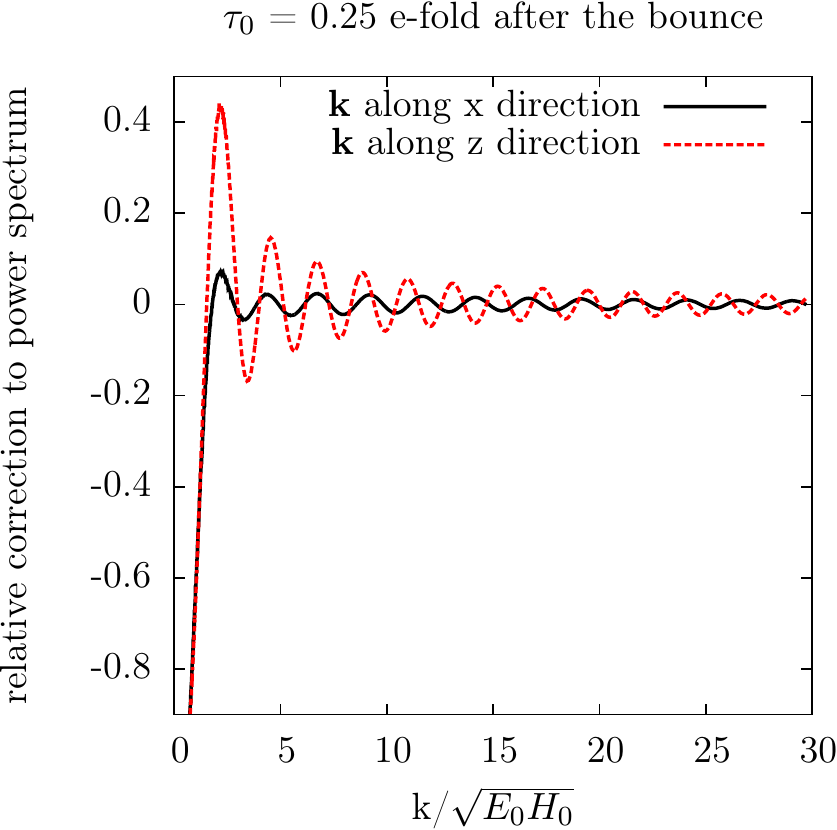}
  \hspace{0.04\textwidth}
  \includegraphics[width=0.47\textwidth]{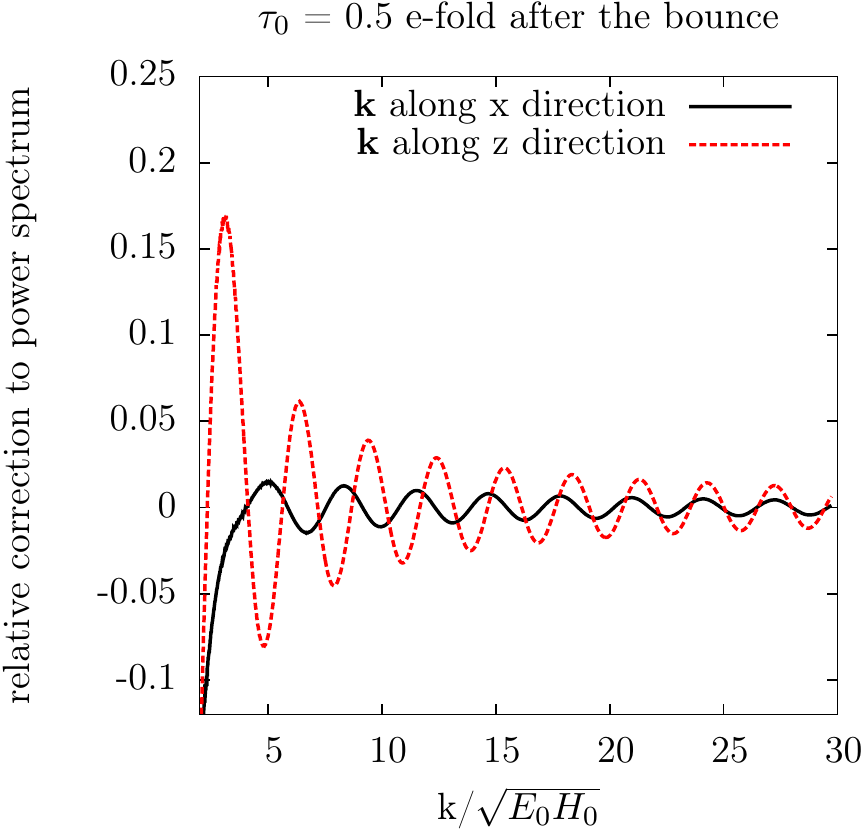}
  \caption{\label{fig:posf}
  If we rescale $\tau \to \tau/\sqrt{E_0 H_0}$ and $k \to k\sqrt{E_0 H_0}$ in Eq.~(\ref{eq:eom-nonpert}), the equation of motion becomes independent of $E_0$ and $H_0$. So we plot $k$ in unit of $\sqrt{E_0 H_0}$.
  In this figure, the initial condition is chosen to be the instant positive frequency at time $\tau_0$. The left and right panels are plotted with initial time $\tau_0$ to be $1/4$ and $1/2$ e-fold after the bounce respectively.}
\end{figure}

\begin{figure}[ht]
  \centering
  \includegraphics[width=0.47\textwidth]{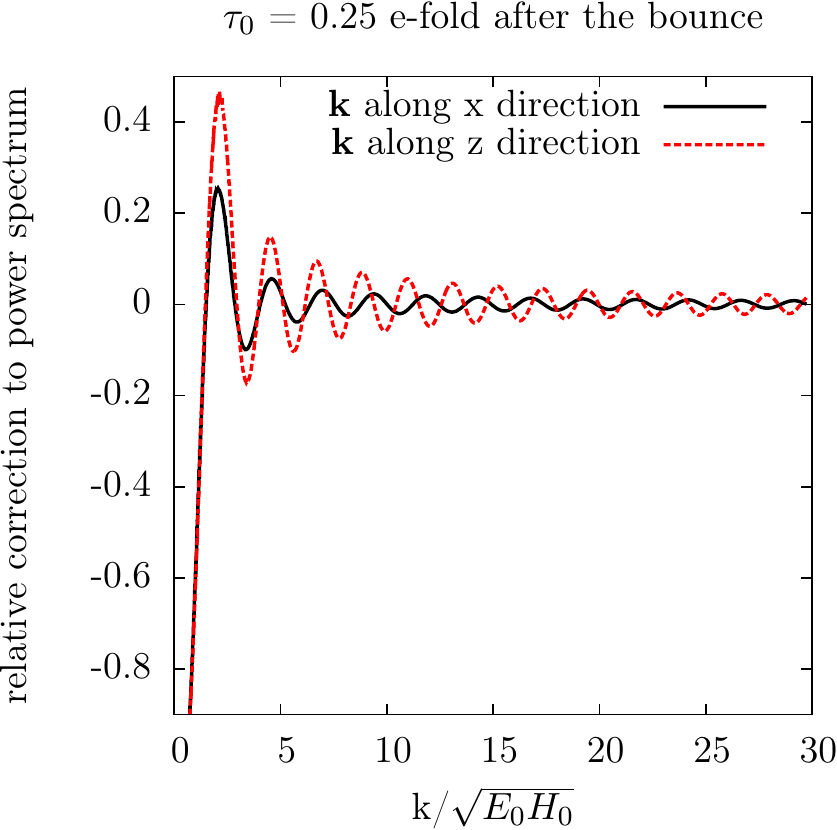}
  \hspace{0.04\textwidth}
  \includegraphics[width=0.47\textwidth]{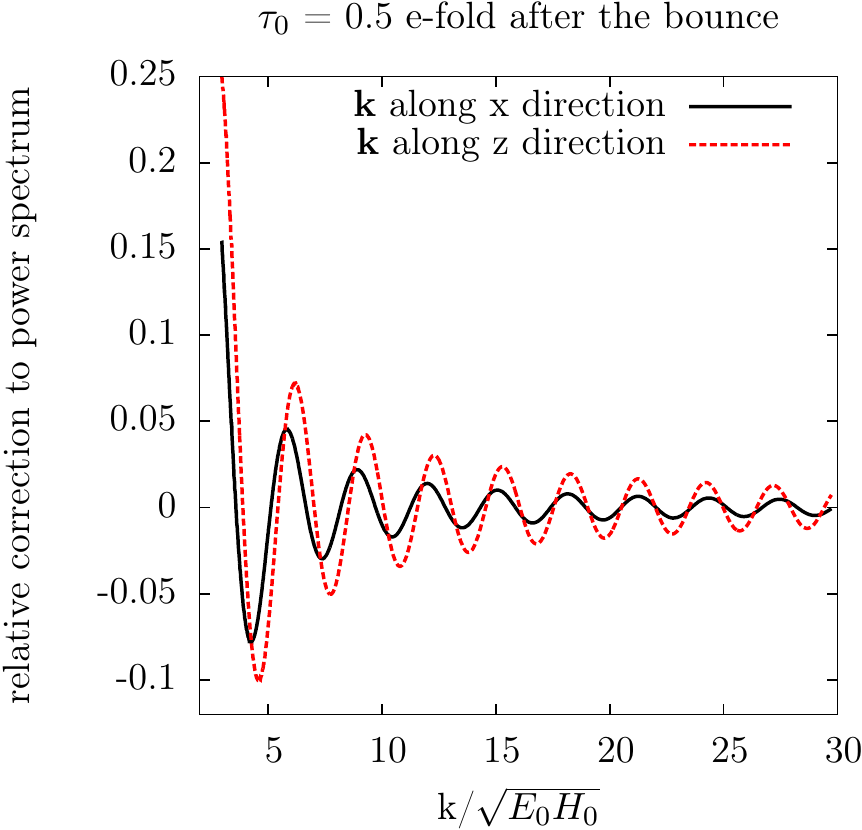}
  \caption{\label{fig:lowe} Similar to Fig.~\ref{fig:posf}, except that the initial condition is chosen to minimize energy at time $\tau_0$.}
\end{figure}

The state with only the BD component is the lowest energy ground state for modes well within the horizon $|k\tau| \ll 1$. This is the reason that the BD vacuum is the typical choice in cases where the inflaton has happened for sufficiently large number of e-folds. But if the inflation only lasts for the minimum number of e-folds, and we are looking at the near-horizon modes at the beginning of the inflation, the BD component is no longer the ground state. This is further complicated by the possibility that, at the beginning of the inflation, we may not even have good reasons to expect that the state would start in the ground state. Both arguments suggest that the non-BD component should be present for these modes generically.

In Fig.~\ref{fig:posf} and Fig.~\ref{fig:lowe} we give two examples of non-BD vacua specified at the initial time $\tau_0$. We compute the power spectrum numerically, which also allows us to solve the equation (\ref{eq:eom-nonpert}) non-perturbatively in $E_0^2$. In the first example, we choose the ``instant positive frequency mode" at $\tau_0$ as the initial condition. This condition is defined by changing the variable $u_\bk$ to $v_\bk$ which satisfies $v_\mathbf{k}'' + k_\mathrm{eff}^2 v_\mathbf{k} = 0$, and requiring $v_\mathbf{k}'=-ik_\mathrm{eff}v_\mathbf{k}$ at $\tau_0$.
Note this is different from the BD component, because in the BD component only the modes well inside the horizon ($|k\tau| \to \infty$) has the ``instant positive frequency" defined in this way.
In the second example, we choose the ``lowest energy state" at $\tau_0$. This state is chosen by numerically sampling different non-BD coefficients and choosing the one which minimizes the Hamiltonian at $\tau_0$. Note that, as mentioned, the lowest energy state at the $|k\tau_0| \gtrsim 1$ is different from the lowest energy state at $|k\tau| \to \infty$ (the BD state).

One important effect that the non-BD component introduces is the sinusoidal scale-dependence.
This generic feature can be seen as follows. We denote the coefficients of the BD and non-BD component as $C_+$ and $C_-$, respectively. Imposing certain relations for the mode function at $\tau_0$ relates $C_+ u_{\bk(0)}$ to $C_- u_{\bk(0)}^*$. So $C_- \propto C_+ e^{-2ik\tau_0}$. The power spectrum is proportional to $|C_+ + C_-|^2$, which then contains a term proportional to $\sin(2k\tau_0 + {\rm phase})$. We will provide more detailed examples in \cite{Toappear}.
This running behavior has the characteristics of those of the sharp feature signal \cite{Chen:2011zf}. Namely, it has the sinusoidal running with a wavelength $\pi/\tau_0$ determined by the location of the feature $\tau_0$.
Another important effect of the non-BD component is that it generically changes the decay behavior of the non-oscillatory component \cite{Toappear}.

\section{Analytical template}
\setcounter{equation}{0}

Now we summarize the main results into the following analytical template for the power spectrum,
\bal
P_\zeta = P_\zeta^{\rm iso} + \alpha \frac{\cos^2\theta}{k^n}
+ \beta \frac{1}{k^m} \sin(2 k \tau_0 + \phi) ~.
\label{P_template}
\eal
\begin{itemize}
\item The first term is the isotropic part of the power spectrum. The relic vector fields can affect this part by introducing some mild scale dependence, $P_\zeta^{\rm iso} = P_{\zeta0} + \tilde\alpha/k^n $, as we have seen in (\ref{power_BD}); but this scale-dependence is typically swamped by the sinusoidal running coming from the third term due to the non-BD component.

\item The second term is the most important anisotropic part introduced by the relic vector field, and it is proportional to $\cos^2\theta$, where $\theta$ is the angle between the comoving momentum $\bk$ and the preferred direction singled out by the vector field. This anisotropic component has a strong scale dependence which decays towards the shorter scales. The details of the decay behavior are model-dependent. Since the massless vector field considered here is diluted as $1/a^4$ as radiation, for the BD case, the $k$-dependence is $\propto 1/k^4$. The non-BD case can change this behavior and make it decay more slowly, e.g.~$n=2$ \cite{Toappear}.

    These properties can potentially explain the quadrupole-octopole alignment. On the one hand, the anisotropic component picks up a preferred direction and align the two (or more) multipoles; on the other hand, such alignment effect decreases towards the larger multipoles.

    This term also predicts off-diagonal terms in the power spectrum $\langle a_{l_1m_1}a^*_{l_2m_2}\rangle$ with specific properties. We show these properties in Fig.~\ref{fig:m0ratio} of Appendix A. These components are valuable because they effectively double the amount of data to be compared with theory; also because they do not have the leading contribution from the isotropic components, and hence provide cleaner signals on the statistical anisotropy that we are probing.

\item The third term is the contribution from the non-BD component, generically present because we are considering the initial moments of inflation.
    At the time $\tau_0$ where we set the initial conditions, the modes observable today are all within or around the horizon crossing.
    If we denote the comoving momentum of the largest observable mode as $k_{\rm CMB}$, then $|k_{\rm CMB}\tau_0| \gtrsim 1$. This means that the sinusoidal running in (\ref{P_template}) has an approximately constant wavelength $\Delta (k/k_{\rm CMB}) \lesssim \CO(1)$. This is one of the most important signatures of the non-BD component. In addition and more model-dependently, this sinusoidal running has an envelop $\sim \beta/k^m$ which decays towards short scales, typically, $m\sim 1$. For simplicity we ignored anisotropic terms in this envelop \cite{Toappear}.
    Due to the envelop, the modulation is larger for the smaller $\ell$, and this potentially can be responsible for the low quadrupole power.
    Whether the quadrupole is anomalously low or high is an accident determined by the phase.

\item Finally, we discuss the effect of this template on the parity anomaly.

\begin{figure}[t]
  \centering
  \includegraphics[width=0.7\textwidth]{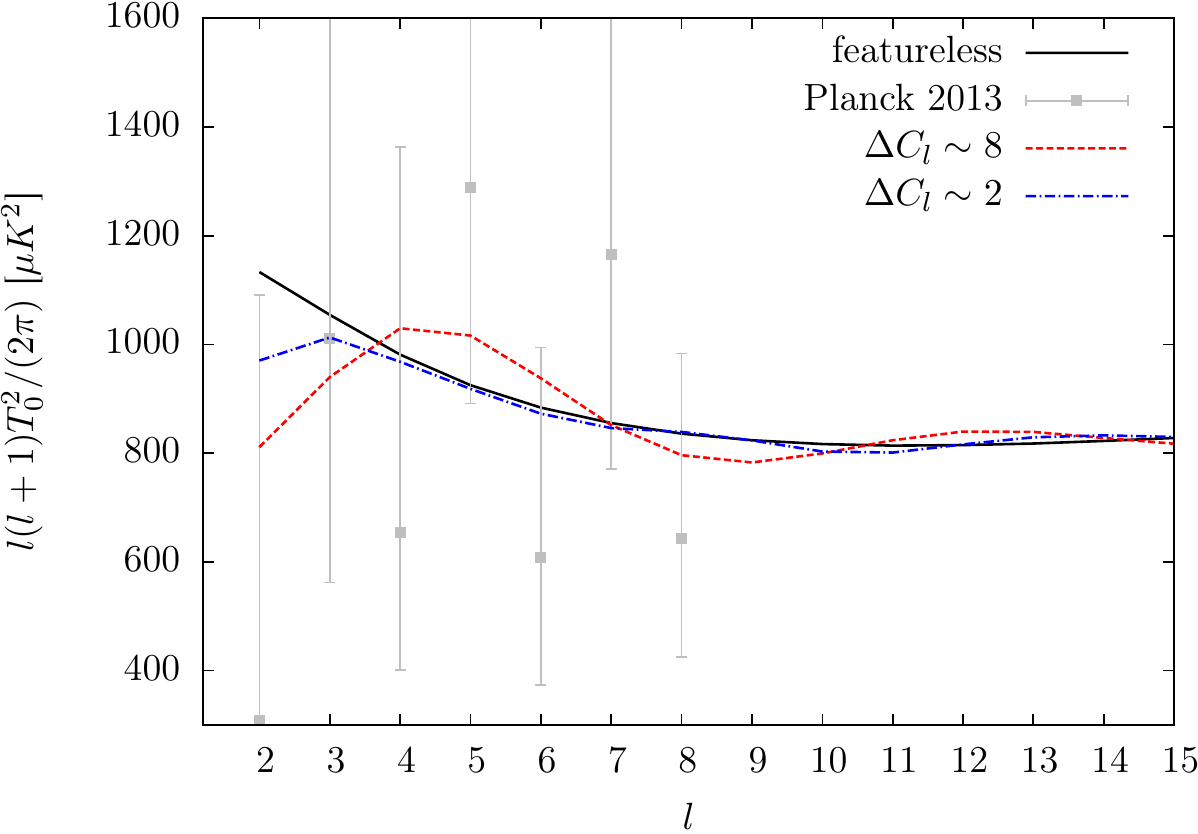}
  \caption{\label{fig:CAMBosc} The impact of the third term in Eq.~\eqref{P_template} on large scale CMB power spectrum. Parameters correspond to those in Fig.~\ref{fig:lowe} ($z$-direction). We choose different $\tau_0$ so that the red dash line and the blue dash-dot line corresponds to the oscillation period $\Delta C_l\sim 8$ and $\Delta C_l\sim 2$, respectively. }
\end{figure}

    As discussed in the Appendix A, the second term in (\ref{P_template}) does not produce $\ell$-dependent effect on the diagonal components of the power spectrum, so we discuss the effect of the isotropic sinusoidal modulation from the third term.
    From the above discussion we know that the underlying primordial oscillation has the periodicity $\Delta(k/k_{\rm CMB}) \lesssim \CO(1)$. This inequality can explain that the observed oscillation has the period $\Delta \ell =2$ in the multipole space.
    Smaller period $\Delta(k/k_{\rm CMB})$ does not change the period $\Delta \ell =2$ which is already the minimum, but reduces the oscillation amplitude in $C_\ell$ due to averaging effect.
    We have tested this approach using the example presented in Fig.~\ref{fig:lowe} using CAMB \cite{Lewis:1999bs}. The result is in Fig.~\ref{fig:CAMBosc}.
    If the sinusoidal modulation has relatively larger period in $\ell$-space, e.g.~$\Delta\ell\gtrsim 4$, the modulation amplitude remains significant. The problem of this approach is that, for $\Delta\ell \approx 2$, the modulation amplitude is significantly reduced due to the projection effect from the $\bk$-space to $\ell$-space, as can be seen from the Sachs-Wolfe approximation. So it cannot explain the oscillation in the parity anomaly in full details. Nonetheless, the suppression at the largest scale can still be significant.

\item So, if taking the typical values $n=2$ (or $4$) and $m=1$, we have four extra parameters: $\alpha$ determines the amplitude of anisotropy; $\beta$ determines the amplitude of the oscillatory modulation; $\tau_0$ determines how fast the underlying primordial oscillation is, and after coarse-graining in the $\ell$ space, modifies the effective envelop of the oscillation; and $\phi$ is the phase of the oscillation.

\end{itemize}

Having concrete models and explicit analytical predictions may provide additional fits that are not obvious if investigated in data analyses alone. So it will be interesting to perform a detailed comparison between the model predictions and data, for example by making connection to the method of \cite{Gordon:2005ai}. It will also be interesting to compare with other possible sources of primordial anisotropy \cite{Pereira:2007yy}, by working out more explicit analytical predictions of these models in the literature using the method in this paper. Such investigations may open up an observational window to the initial moments of the primordial inflation.

Finally, we have studied the statistical anisotropies on the temperature map of the CMB. It would be interesting to extend the investigation to E and B-mode polarizations. Similar anisotropies and corrections to the power spectrum should show up in the TE and EE correlations because their sources are dominated by the inflaton fluctuations. On the other hand, the B-mode polarization, originated from the primordial gravitational waves, deserves a separate study, including the possible difference in the initial conditions and in the dynamics of the perturbations.

\medskip
\section*{Acknowledgments}

We would like to thank James Fergusson for helpful discussions, and Simon Su for valuable help on the CMB analyses.
XC and YW are supported in part by the Stephen Hawking Advanced Fellowship. YW is supported in part by fundings from Kavli IPMU (WPI), the University of Tokyo, and a Starting Grant of the European Research Council
(ERC STG grant 279617).

\appendix

\section{Anisotropy projected onto the CMB}

Here we relate the primordial anisotropy to the CMB anisotropies. For construction of estimators and Fisher matrix analysis we refer to \cite{Ma:2011ii}. $C_l$ and $a_{lm}$ can be calculated as
\begin{align} \label{eq:alm}
  a_{lm} = 4\pi (-i)^l \int \frac{d^3 k}{(2\pi)^3} ~ g_l (k) \zeta_\mathbf{k} Y^*_{lm}(\hat k)~, \qquad
  C_l = \frac{1}{2l+1} \sum_m \langle a_{lm}a^*_{lm}\rangle ~.
\end{align}
where $g_l(k)$ is the radiation transfer function.

Inserting the anisotropic term with $\cos^2\theta$ anisotropy, the correction to $C_l$ now takes the form
\begin{align}
  \Delta C_l =
  \left[ 4\pi \alpha \int dk ~ \frac{g_l^2(k)}{k^{n+1}}  \right]
  \left[ \frac{1}{2l+1}  \sum_m \int d\Omega_k ~ Y_{lm}(\mathbf{k}) Y^*_{lm}(\mathbf{k}) \cos^2\theta\right]~.
\end{align}
Here the first $\left[ \cdots \right]$ can be calculated in CAMB, as if the power spectrum is $\Delta P(k)= \alpha/k^n$. This is because if we had set $\cos^2\theta =1$, the second $\left[ \cdots \right]$ is one. On the other hand, the second $\left[ \cdots \right]$ is
\begin{align} \label{eq:aniso-integ}
  \frac{1}{2l+1}  \sum_m \int d\Omega_k ~ Y_{lm}(\mathbf{k}) Y^*_{lm}(\mathbf{k}) \cos^2\theta = \frac{1}{3} ~.
\end{align}
This integral is $l$-independent. In other words, the anisotropy does not help for resolving the parity anomaly \footnote{On the other hand, if the anisotropy was of the dipolar type (i.e. replacing the $\cos^2\theta$ with $\cos\theta$ in Eq. \eqref{eq:aniso-integ}), the anisotropy does not help for the parity anomaly either because the integral in \eqref{eq:aniso-integ} then becomes zero.}.

\begin{figure}[htbp]
  \centering
  \includegraphics[width=0.7\textwidth]{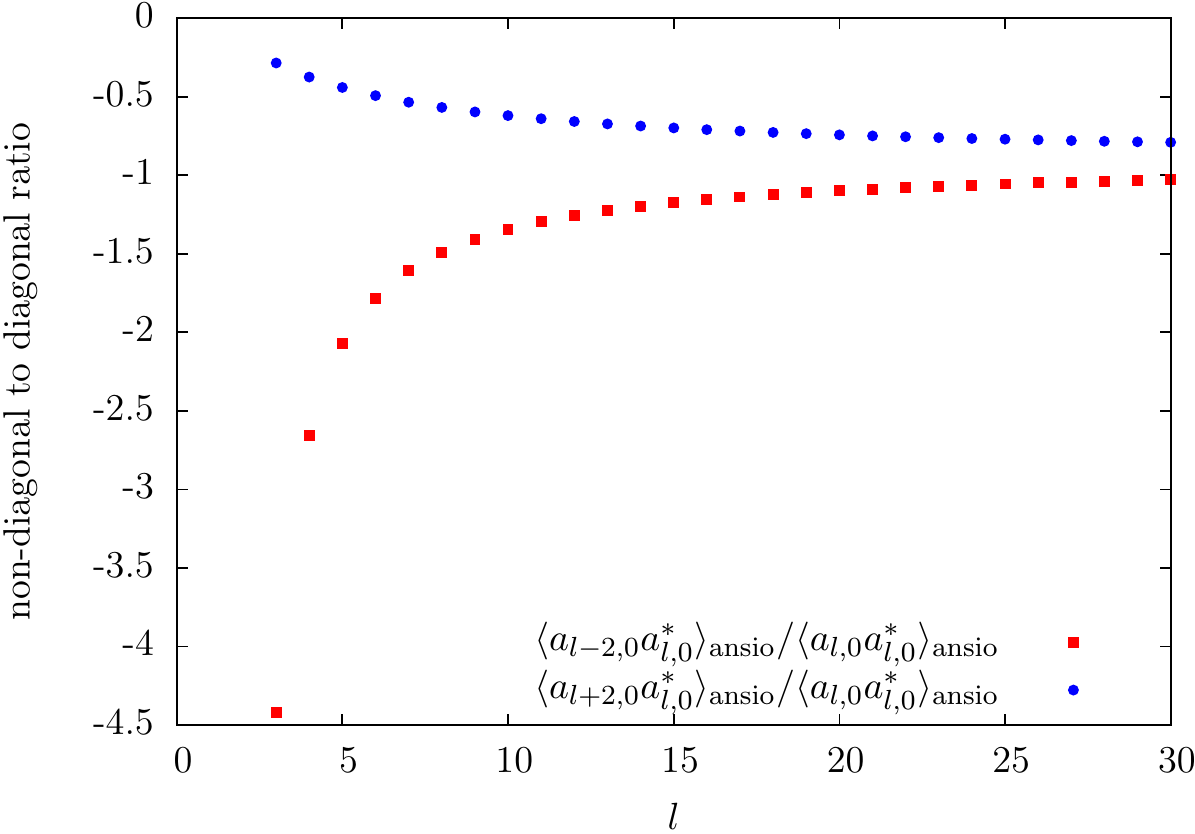}
  \caption{\label{fig:m0ratio} Relative size of the non-diagonal and diagonal correlators. Here in the transfer function part, Sachs-Wolfe approximation is used. We have chosen $m_1=m_2=0$. The behavior of non-zero $m_1, m_2$ are tested to behave similarly. }
\end{figure}

On the other hand, it is more interesting to consider correlators with different multiple moment. In this case, using the Gaunt's formula
\begin{align}
  & \int_0^\pi  d\theta \int_0^{2\pi} d\phi  ~ Y^*_{l_1m_1}Y_{l_2m_2} Y_{l_3m_3}
  \nonumber\\ =& ~
  (-1)^{m_1} \sqrt{\frac{(2l_1+1)(2l_2+1)(2l_3+1)}{4\pi}}
  \left( \begin{array}{ccc}
      l_1 & l_2 & l_3\\
      0 & 0 & 0
    \end{array} \right)
  \left( \begin{array}{ccc}
      l_1 & l_2 & l_3\\
      -m_1 & m_2 & m_3
    \end{array} \right) ~,
\end{align}
and the spherical harmonics representation of triangle function
\begin{align}
  \cos^2 \theta = \frac{1}{3} \left( 4 \sqrt{\frac{\pi}{5} } Y_{2,0}(\theta,\phi)+1\right)~,
\end{align}
The correlator can be calculated as
\begin{align} \label{eq:lmlmc}
  & \langle a_{l_1m_1}a^*_{l_2m_2}\rangle_\mathrm{ansio}
  \nonumber\\
  = &~
  \alpha \frac{8\pi}{3} i^{l_2-l_1}(-1)^{m_1} \sqrt{(2l_1+1)(2l_2+1)}
  \left( \begin{array}{ccc}
      l_1 & l_2 & 2\\
      0 & 0 & 0
    \end{array} \right)
  \left( \begin{array}{ccc}
      l_1 & l_2 & 2\\
      -m_1 & m_2 & 0
    \end{array} \right)  \int \frac{dk}{k^{n+1}}~ g_{l_1}(k)g_{l_2}(k)~,
\end{align}
where the subscript aniso denotes the part proportional to $Y_{20}$. Here we have used the Wigner's 3j symbol. Note that RHS of \eqref{eq:lmlmc} is non-vanishing only when
\begin{itemize}
\item $m_1=m_2$.
\item  Either $l_1=l_2$ or $l_1 = l_2 \pm 2$.  This is because the 3j symbol satisfies triangle identity $l_2 - 2 \leq  l_1 \leq l_2 + 2$. Thus for non-vanishing $\langle a_{l_1m_1}a^*_{l_2m_2}\rangle_\mathrm{ansio} $, we need $l_1 = l_2 \pm n$, where $n=0,1,2$. However applying the symmetry of the 3j symbol
\begin{align}
    \left( \begin{array}{ccc}
      l_1 & l_2 & l_3\\
      m_1 & m_2 & m_3
    \end{array} \right)
  =
    (-1)^{l_1+l_2+l_3} \left( \begin{array}{ccc}
      l_1 & l_2 & l_3\\
      -m_1 & -m_2 & -m_3
    \end{array} \right)
\end{align}
to
\begin{align}
  \left( \begin{array}{ccc}
      l_1 & l_2 & 2\\
      0 & 0 & 0
    \end{array} \right) ~,
\end{align}
the possibility $l_1=l_2\pm 1$ is excluded.
\item Numerically, the size of the non-diagonal and diagonal correlators are comparable. For example, for $n=2$ and $m_1=m_2=0$, the relative size are plotted in Fig. \ref{fig:m0ratio}. Different choices of $n$ and $m$ yields similar result.
\end{itemize}

Note that the off-diagonal (different $l$) components double the amount of data for the purpose of analysis. Thus taking those off-diagonal components into consideration reduces the error bar by about $1/\sqrt{2}$. This is especially valuable considering the cosmic variance at large scales. Moreover, considering that the off-diagonal components have no isotropic part, the message in the off-diagonal components should be cleaner than the one in the diagonal components.

The integral $\int \frac{dk}{k^{n+1}}~ g_{l_1}(k)g_{l_2}(k)$ can be in general calculated numerically. In the Sachs-Wolfe limit, $g_l(k) = - j_l(k r_\mathrm{rec})/5$ and the integral takes the form
\begin{align}
\int \frac{dk}{k^{n+1}}~ g_{l_1}(k)g_{l_2}(k) = &
  \frac{\pi  r_\mathrm{rec}^n}{25 \times  2^{n+3}} \frac{\Gamma((l_1+l_2-n)/2)}{\Gamma((3+l_1-l_2+n)/2)\Gamma((3+2l_2)/2)}
\nonumber\\  & \times {}_2F_1 ((-1-l_1+l_2-n)/2, (l_1+l_2-n)/2, (3+2l_2)/2, 1)~,
\end{align}
and ${}_2F_1(a,b,c,z)$ is the hypergeometric function.

\end{spacing}

\end{document}